\documentclass[zpreprint,zbstdefault]{./LaTeX/zeus/zeus_paper}
\usepackage[english]{babel}

\newcommand{\ZcoosysB}{%
The ZEUS coordinate system is a right-handed Cartesian system, with the $Z$
axis pointing in the proton beam direction, referred to as the ``forward
direction'', and the $X$ axis pointing left towards the centre of HERA.
The coordinate origin is at the nominal interaction point.\xspace}
\newcommand{\Zpsrap}{%
The pseudorapidity is defined as $\eta=-\ln\left(\tan\frac{\theta}{2}\right)$,
where the polar angle, $\theta$, is measured with respect to the proton beam
direction.\xspace}

\newcommand{\ZcoosysfnBeta}{\footnote{\ZcoosysB\Zpsrap}}
\newcommand{\Zdetdesc}{%
A detailed description of the ZEUS detector can be found 
elsewhere~\cite{zeus:1993:bluebook}. A brief outline of the 
components that are most relevant for this analysis is given
below.\xspace}
\newcommand{\Zctddesc}[1]{%
Charged particles are tracked in the central tracking detector (CTD)~\citeCTD,
which operates in a magnetic field of $1.43\Tesla$ provided by a thin 
superconducting coil. The CTD consists of 72~cylindrical drift chamber 
layers, organized in 9~superlayers covering the polar-angle#1 region 
\mbox{$15^\circ<\theta<164^\circ$}. The transverse-momentum resolution for
full-length tracks is $\sigma(p_T)/p_T=0.0058p_T\oplus0.0065\oplus0.0014/p_T$,
with $p_T$ in $\Gev$.}
\newcommand{\Zcaldesc}{%
The high-resolution uranium--scintillator calorimeter (CAL)~\citeCAL consists 
of three parts: the forward (FCAL), the barrel (BCAL) and the rear (RCAL)
calorimeters. Each part is subdivided transversely into towers and
longitudinally into one electromagnetic section (EMC) and either one (in RCAL)
or two (in BCAL and FCAL) hadronic sections (HAC). The smallest subdivision of
the calorimeter is called a cell.  The CAL energy resolutions, as measured under
test-beam conditions, are $\sigma(E)/E=0.18/\sqrt{E}$ for electrons and
$\sigma(E)/E=0.35/\sqrt{E}$ for hadrons ($E$ in $\Gev$).}

\chardef\usc=95
\chardef\til=126
\catcode`\@=11 
\DeclareRobustCommand\xdotspace{\futurelet\@let@token\@xdotspace}
\def\@xdotspace{%
  \ifx\@let@token.\else
  \ifx\@let@token\bgroup.\else
  \ifx\@let@token\egroup.\else
  \ifx\@let@token\/.\else
  \ifx\@let@token\ .\else
  \ifx\@let@token~.\else
  \ifx\@let@token!.\else
  \ifx\@let@token,.\else
  \ifx\@let@token:.\else
  \ifx\@let@token;.\else
  \ifx\@let@token?.\else
  \ifx\@let@token/.\else
  \ifx\@let@token'.\else
  \ifx\@let@token).\else
  \ifx\@let@token-.\else
  \ifx\@let@token\@xobeysp.\else
  \ifx\@let@token\space.\else
  \ifx\@let@token\@sptoken.\else
   .\space
   \fi\fi\fi\fi\fi\fi\fi\fi\fi\fi\fi\fi\fi\fi\fi\fi\fi\fi}
\catcode`\@=12 

\newcommand{\stru}[2]{%
   \relax\ifmmode\hbox{\vrule height#1 depth#2 width0pt}%
   \else\vrule height#1 depth#2 width0pt\fi}

\newcommand{\Ronum}[1]{\uppercase\expandafter{\romannumeral#1}}
\newcommand{\ronum}[1]{\expandafter{\romannumeral#1}}
\DeclareRobustCommand{\LaTeXZ}{%
  \LaTeX\kern-.05em4\kern-.1em
  {\raisebox{-0.2ex}{$\scriptstyle\text{ZEUS}$}}\xspace}

\newcommand{\fig}[1]{Fig.~\ref{fig-#1}}
\newcommand{\Fig}[1]{Figure~\ref{fig-#1}}

\newcommand{\tab}[1]{Table~\ref{tab-#1}}
\newcommand{\Tab}[1]{Table~\ref{tab-#1}}

\newcommand{\sect}[1]{Sect.~\ref{sec-#1}}
\newcommand{\Sect}[1]{Section~\ref{sec-#1}}

\DeclareMathAlphabet{\mathbf}{OT1}{cmr}{bx}{sl}
\newcommand{\eVdist}{\kern-0.06667em}

\newcommand{\Gev}{{\text{Ge}\eVdist\text{V\/}}}

\newcommand{\mev}{{\,\text{Me}\eVdist\text{V\/}}}
\newcommand{\gev}{{\,\text{Ge}\eVdist\text{V\/}}}

\newcommand{\nb}{\,\text{nb}}
\newcommand{\pb}{\,\text{pb}}

\newcommand{\pbi}{\,\text{pb}^{-1}}

\newcommand{\met}{\,\text{m}}

\newcommand{\Tesla}{\,\text{T}}

\newcommand{\slashfrac}[2]{%
  \raisebox{0.5ex}{\ensuremath #1}\kern-0.12em/\kern-0.08em
  \raisebox{-.8ex}{\ensuremath #2}}

\newcommand{\sqr}[3]{%
    {\vcenter{\hrule height.#3ex\hbox{\vrule width.#2ex height#1ex
     \kern#1ex\vrule width.#3ex}\hrule height.#2ex}}}

\newcommand{\widebar}[1]{%
   \mkern1.5mu\overline{\mkern-1.5mu#1\mkern-1.mu}\mkern1.mu}
\catcode`\@=11 
\newcommand{\parenbar}{\mathpalette\p@renb@r}
\def\p@renb@r#1#2{\vbox{%
  \ifx#1\scriptscriptstyle \dimen@.7em\dimen@ii.2em\else
  \ifx#1\scriptstyle \dimen@.8em\dimen@ii.25em\else
  \dimen@1em\dimen@ii.4em\fi\fi \offinterlineskip
  \ialign{\hfill##\hfill\cr
    \vbox{\hrule width\dimen@ii}\cr
    \noalign{\vskip-.3ex}%
    \hbox to\dimen@{$\mathchar300\hfil\mathchar301$}\cr
    \noalign{\vskip-.3ex}%
    $#1#2$\cr}}}
\catcode`\@=12 

\newcommand{\qbar}{\widebar{q}}

\newcommand{\diff}{{\rm d}}

\newcommand{\IP}{{\rm I$\kern-0.01667em$P}\xspace}

\newcommand{\smax}{{\rm max}}

\newcommand{\Lumi}{{\cal L}}

\mathchardef\qsm=63
\mathchardef\pls=43
\mathchardef\mns=512
\mathchardef\plm=518
\mathchardef\eql=61
\mathchardef\smallleft=300
\mathchardef\smallright=301
\mathchardef\les=316
\mathchardef\gre=318
\mathchardef\leq=532
\mathchardef\grq=533

\catcode`\@=11 
\newcounter{pict@width}
\newcounter{pict@height}
\newlength{\pict@scale}
\newcommand{\psfigadd}[4]{%
\setcounter{pict@width}{1*\ratio{#2+\pict@scale/2}{\pict@scale}}
\setcounter{pict@height}{1*\ratio{#3+\pict@scale/2}{\pict@scale}}
\setlength{\unitlength}{\pict@scale}
\hbox to #2{\hspace{-\fill}\begin{picture}(\thepict@width,\thepict@height)
\put(0,0){\psfig{figure=#1,width=#2,height=#3,clip=}}
\SetScale{0.283466457}
\SetWidth{1.763889}
{#4}
\end{picture}}
}
\newcounter{pict@widthfst}
\newcounter{pict@widthscd}
\newcounter{pict@widthtot}
\newcommand{\psfigaddtwo}[7]{%
\setcounter{pict@widthfst}{1*\ratio{#2+\pict@scale/2}{\pict@scale}}
\setcounter{pict@widthscd}{1*\ratio{#2+#4+\pict@scale/2}{\pict@scale}}
\setcounter{pict@widthtot}{1*\ratio{#2+#4+#6+\pict@scale/2}{\pict@scale}}
\setcounter{pict@height}{1*\ratio{#3+\pict@scale/2}{\pict@scale}}
\setlength{\unitlength}{\pict@scale}
\hbox{\hspace{-\fill}\begin{picture}(\thepict@widthtot,\thepict@height)
\put(0,0){\psfig{figure=#1,width=#2,height=#3,clip=}}
\put(\thepict@widthscd,0){\psfig{figure=#5,width=#6,height=#3,clip=}}
\SetScale{0.283466457}
\SetWidth{1.763889}
{#7}
\end{picture}}
}
\newcommand{\psfigror}[4]{%
\setcounter{pict@width}{1*\ratio{#2+\pict@scale/2}{\pict@scale}}
\setcounter{pict@height}{1*\ratio{#3+\pict@scale/2}{\pict@scale}}
\setlength{\unitlength}{\pict@scale}
\hbox{\begin{picture}(\thepict@width,\thepict@height)
\put(0,\thepict@height){\psfig{figure=#1,width=#3,height=#2,clip=,angle=270}}
\SetScale{0.283466457}
\SetWidth{1.763889}
{#4}
\end{picture}}
}
\newcommand{\psfigrol}[4]{%
\setcounter{pict@width}{1*\ratio{#2+\pict@scale/2}{\pict@scale}}
\setcounter{pict@height}{1*\ratio{#3+\pict@scale/2}{\pict@scale}}
\setlength{\unitlength}{\pict@scale}
\hbox{\begin{picture}(\thepict@width,\thepict@height)
\put(0,0){\psfig{figure=#1,width=#3,height=#2,clip=,angle=90}}
\SetScale{0.283466457}
\SetWidth{1.763889}
{#4}
\end{picture}}
}
\catcode`\@=12 
\newlength\listtextwidth

\catcode`\@=11 
\newlength{\@tabfninsert}
\newlength{\@tabfnwidth}
\newcommand{\tabfootnote}[2]{%
  \setlength{\@tabfninsert}{0.8em}
  \setlength{\@tabfnwidth}{\textwidth}
  \addtolength{\@tabfnwidth}{-\@tabfninsert}
  \addtolength{\@tabfnwidth}{-0.4em}
  \noindent\makebox[\@tabfninsert][r]{\footnotesize$^{#1}$\hfil}\hfill%
  \parbox[t]{\@tabfnwidth}{\footnotesize #2\hfill}}
\catcode`\@=12 

\newcommand {\pom} {I\!\!P}
\newcommand {\pomsub} {{\scriptscriptstyle \pom}}

\newcommand {\xpom} {x_{\pomsub}}
\newcommand {\apom} {\alpha_{\pomsub}}

\newcommand {\qqb} {q \qbar}
\newcommand {\dstar} {D^{\ast \pm}}
\newcommand {\etam} {\eta_\smax}

\def\citeCTD{{\cite{%
nim:a279:290,*npps:b32:181,*nim:a338:254%
}}\xspace}
\def\citeCAL{{\cite{%
nim:a309:77,*nim:a309:101,*nim:a321:356,*nim:a336:23%
}}\xspace}

\begin{document}
\prepnum{DESY--02--082}

\title{
Measurement of diffractive production of \\
$\mathbf{\dstar}$(2010) mesons in deep inelastic \\
scattering at HERA \\
}                                                       
                    
\author{ZEUS Collaboration}
\date{4.\ June 2002}

\abstract{
Diffractive production of $\dstar (2010)$ mesons in deep inelastic 
scattering has been measured with the ZEUS detector at HERA using an 
integrated luminosity of $44.3\pbi$.  Diffractive charm production is 
identified by the presence of a large rapidity gap in the final state of 
events in which a $\dstar (2010)$ meson is reconstructed in the decay 
channel 
$D^{* +} \rightarrow (D^0 \rightarrow K^-\pi^+)\pi^+_s $ 
(+ charge conjugate).  Differential cross sections when compared with 
theoretical predictions indicate the importance of gluons in such diffractive
interactions.
}

\makezeustitle

%
%
%
%
\pagenumbering{Roman}                                                             
\begin{center}                                                                   
{                      \Large  The ZEUS Collaboration              }            
\end{center}                                                                                      
  S.~Chekanov,                                                                                     
  D.~Krakauer,                                                                                     
  S.~Magill,                                                                                       
  B.~Musgrave,                                                                                     
  J.~Repond,                                                                                       
  R.~Yoshida\\                                                                                     
 {\it Argonne National Laboratory, Argonne, Illinois 60439-4815}~$^{n}$                            
\par \filbreak                                                                                     
  M.C.K.~Mattingly \\                                                                              
 {\it Andrews University, Berrien Springs, Michigan 49104-0380}                                    
\par \filbreak                                                                                     
  P.~Antonioli,                                                                                    
  G.~Bari,                                                                                         
  M.~Basile,                                                                                       
  L.~Bellagamba,                                                                                   
  D.~Boscherini,                                                                                   
  A.~Bruni,                                                                                        
  G.~Bruni,                                                                                        
  G.~Cara~Romeo,                                                                                   
  L.~Cifarelli,                                                                                    
  F.~Cindolo,                                                                                      
  A.~Contin,                                                                                       
  M.~Corradi,                                                                                      
  S.~De~Pasquale,                                                                                  
  P.~Giusti,                                                                                       
  G.~Iacobucci,                                                                                    
  A.~Margotti,                                                                                     
  R.~Nania,                                                                                        
  F.~Palmonari,                                                                                    
  A.~Pesci,                                                                                        
  G.~Sartorelli,                                                                                   
  A.~Zichichi  \\                                                                                  
  {\it University and INFN Bologna, Bologna, Italy}~$^{e}$
% ...
\newpage                                                                                           
% ...
                                                           %
                                                           %
\newpage   
                                                           %
                                                           %
\begin{tabular}[h]{rp{14cm}}                                                                       
$^{a}$ &  supported by the Natural Sciences and Engineering Research                               
          Council of Canada (NSERC) \\                                                             
$^{b}$ &  supported by the German Federal Ministry for Education and                               
          Research (BMBF), under contract numbers HZ1GUA 2, HZ1GUB 0, HZ1PDA 5, HZ1VFA 5\\         
$^{c}$ &  supported by the MINERVA Gesellschaft f\"ur Forschung GmbH, the                          
          Israel Science Foundation, the U.S.-Israel Binational Science                            
          Foundation, the Israel Ministry of Science and the Benozyio Center                       
          for High Energy Physics\\                                                                
$^{d}$ &  supported by the German-Israeli Foundation, the Israel Science                           
          Foundation, and by the Israel Ministry of Science\\                                      
$^{e}$ &  supported by the Italian National Institute for Nuclear Physics (INFN) \\                
$^{f}$ &  supported by the Japanese Ministry of Education, Science and                             
          Culture (the Monbusho) and its grants for Scientific Research\\                          
$^{g}$ &  supported by the Korean Ministry of Education and Korea Science                          
          and Engineering Foundation\\                                                             
$^{h}$ &  supported by the Netherlands Foundation for Research on Matter (FOM)\\                   
$^{i}$ &  supported by the Polish State Committee for Scientific Research,                         
          grant no. 620/E-77/SPUB-M/DESY/P-03/DZ 247/2000-2002\\                                   
$^{j}$ &  partially supported by the German Federal Ministry for Education                         
          and Research (BMBF)\\                                                                    
$^{k}$ &  supported by the Fund for Fundamental Research of Russian Ministry                       
          for Science and Edu\-cation and by the German Federal Ministry for                       
          Education and Research (BMBF)\\                                                          
$^{l}$ &  supported by the Spanish Ministry of Education and Science                               
          through funds provided by CICYT\\                                                        
$^{m}$ &  supported by the Particle Physics and Astronomy Research Council, UK\\                   
$^{n}$ &  supported by the US Department of Energy\\                                               
$^{o}$ &  supported by the US National Science Foundation\\                                        
$^{p}$ &  supported by the Polish State Committee for Scientific Research,                         
          grant no. 112/E-356/SPUB-M/DESY/P-03/DZ 301/2000-2002, 2 P03B 13922\\                    
$^{q}$ &  supported by the Polish State Committee for Scientific Research,                         
          grant no. 115/E-343/SPUB-M/DESY/P-03/DZ 121/2001-2002, 2 P03B 07022\\                    
\end{tabular}                                                                                      
                                                           %
\clearpage                                                           %

\pagenumbering{arabic} 
\pagestyle{plain}
\section{Introduction}
\label{sec-int}

Diffractive interactions in neutral current deep inelastic scattering (DIS) 
have been studied extensively at HERA \cite{zfp:c68:569,pl:b348:681,zfp:c70:391,epj:c6:43,pl:b428:206,zfp:c76:613,pr:d65:052001}.  Both inclusive and 
diffractive DIS cross sections rise more rapidly with energy than is the case 
for soft hadronic interactions~\cite{np:b231:189,*pl:b296:227}, indicating the 
presence of a hard process to which perturbative Quantum Chromodynamics (pQCD)
is applicable.  Charm production is a key process for investigating the 
dynamics of diffractive DIS~\cite{pl:b404:353,epj:c1:293,zfp:c74:671,epj:c1:547,pl:b378:347,pl:b406:171,*epj:c11:111,*hep-ph-0010300},
since the charm-quark mass provides a hard scale and charm production is known
to be sensitive to gluon-exchange processes in DIS~\cite{epj:c12:35,np:b545:21}.

Two contrasting approaches to describe diffractive DIS are considered in this
paper:
\begin{enumerate}
\item {\bf Resolved-Pomeron models:} \newline
these are Regge-inspired, with an exchanged Pomeron having a partonic 
structure~\cite{pl:b152:256}.  The evolution of the parton distributions of the
Pomeron with $Q^2$ is described by the DGLAP equations
\cite{sovjnp:15:438,*sovjnp:20:94,*jetp:46:641,*np:b126:298}.  Diffractive HERA
data \cite{pl:b356:129,zfp:c76:613} indicate a large gluonic component.  In
these models, charm is produced via boson-gluon fusion (BGF);
\item {\bf Two-gluon-exchange models:} \newline
these are based on decomposing the wave-function of the virtual photon in the
proton rest frame into partonic Fock states --- particularly $\qqb$ and 
$\qqb g$\cite{QCDmodels}.  These states then interact with the proton via 
colour-singlet exchange, the simplest form of which is the exchange of two 
gluons \cite{pr:d12:163,*prl:34:1286,*pr:d14:246}.  If the $\qqb$ state 
dominates, charm production will be suppressed \cite{zfp:c53:331}.  However, if
the $\qqb g$ state is dominant, similar production rates to those predicted by
the resolved Pomeron model are expected~\cite{jetp:81:625,pl:b378:347,zfp:c74:671}.  The $\qqb$ and $\qqb g$ configurations populate different regions of 
phase space.
\end{enumerate}

The ``soft colour interaction''~\cite{pl:b366:371} and 
``semi-classical''~\cite{pl:b404:353} models give very 
similar predictions for charm production to the two models described 
above and therefore are not considered separately.

This paper describes the measurement of differential cross sections for 
$\dstar$ production in diffractive DIS and the ratio with inclusive DIS 
$\dstar$ production.  The results are compared to resolved-Pomeron and 
two-gluon-exchange models.  Similar results, including comparisons to the 
soft colour interaction, semi-classical and other models, have recently been 
published by the H1 Collaboration \cite{pl:b520:191}.

\section{Description of the experiment}
\label{sec-exp}

The integrated luminosity of $44.3\pm 0.7\pbi$ used for this measurement was
collected at the $ep$ collider HERA with the ZEUS detector during 1995 - 
1997, when HERA collided $27.5\gev$ positrons with $820\gev$ protons, giving a
centre-of-mass energy of $300\gev$.

\Zdetdesc
\Zctddesc{\ZcoosysfnBeta}
\Zcaldesc

The position of positrons scattered with a small angle with respect to the 
positron beam direction was measured using the small-angle rear tracking 
detector (SRTD)~\cite{nim:a401:63}, which also provides a means of correcting 
for any energy loss of the scattered positron due to the presence of inactive 
material.  Complementing the SRTD is the rear presampler 
(RPRES)~\cite{nim:a382:419}, which provides energy-loss information in RCAL 
regions outside the acceptance of the SRTD.

The luminosity was determined from the rate of the bremsstrahlung process
$ep \rightarrow e \gamma p$, where the photon was measured with a 
lead-scintillator calorimeter~\cite{desy-92-066,*zfp:c63:391,*acpp:b32:2025} 
located at $Z = -107\met$.

\section{Kinematics of diffractive DIS}

The kinematics of the inclusive deep inelastic scattering of positrons and 
protons are specified by the positron-proton centre-of-mass energy, 
$\sqrt{s}$, and any two of the following variables:  $Q^2$, the negative
square of the four-momentum of the exchanged photon; $y$, the inelasticity;
$x$, the Bjorken scaling variable and $W$, the centre-of-mass energy of the
photon-proton system.

Additional variables are required to describe the diffractive process
$ep \rightarrow eXp$, where $X \rightarrow \dstar X^{\prime}$.  These 
are:

\begin{itemize}
\item $t$, the square of the four-momentum transfer at the proton vertex.  
Since $t$ was not measured for the present data, all results discussed here are
integrated over this variable;
\item $\xpom = (M_{X}^{2} + Q^2)/(W^2 + Q^2)$, where $M_X$ is the invariant 
mass of the hadronic final state, $X$, into which the virtual photon 
dissociates.  This variable is the fraction of the proton's momentum carried 
by the exchanged colour-singlet system;
\item $\beta = Q^2/(M_{X}^{2} + Q^2)$ can be interpreted within the 
resolved-Pomeron model as the fraction of the Pomeron's momentum carried 
by the struck parton.
\end{itemize}

The above formulae for $\xpom$ and $\beta$ neglect the proton mass and assume 
$t = 0$.  The variables are related by $x = \xpom \beta$.

To reconstruct the kinematic variables, both the final-state positron and the 
hadronic final state must be measured.  The positron was identified
using an algorithm based on a neural network \cite{nim:a365:508}.  The hadronic
final state was reconstructed using combinations of calorimeter cells and CTD 
tracks to form energy-flow objects (EFOs)
\cite{epj:c6:43,briskin:phd:1998,*epj:c1:81}.  The DIS variables were 
reconstructed using the double angle (DA) method 
\cite{proc:hera:1991:23,*proc:hera:1991:43}.

The mass of the diffractive system $X$ was calculated from the EFOs as
\begin{eqnarray}
M_{X}^{2} = \left( \sum_i E_i \right)^2 - \left( \sum_i P_{x_i} \right)^2  - \left( \sum_i P_{y_i} \right)^2  - \left( \sum_i P_{z_i} \right)^2, \nonumber
\end{eqnarray}

where the sum runs over all the EFOs in the event, excluding those 
associated with the scattered positron.

\section{Event Selection}
\subsection{DIS selection and $\mathbf{\dstar \rm(2010)}$ reconstruction}

The initial event sample was selected by identifying DIS events.  The 
selection, both online and offline, was performed in the same manner
as in the inclusive DIS $\dstar$ study~\cite{epj:c12:35}.  The only difference 
is in the kinematic region used, which for the present analysis is 
$4 < Q^2 < 400\gev^2$ and $0.02 < y < 0.7$.

The $\dstar$ selection cuts applied to reduce the combinatorial background 
differ somewhat from those used in the inclusive measurement,
although the analysis is based on the same decay channel: 
$D^{* +} \rightarrow (D^0 \rightarrow K^- \pi^+) \pi^+_s$ (+ charge 
conjugate), where $\pi_s$ indicates the ``slow'' pion~\cite{prl:35:1672}.  
Reduction of the combinatorial background was achieved by requiring:
\begin{itemize}
\item the transverse momenta of any two oppositely charged tracks, assumed 
to be the $K$ and the $\pi$ from the $D^0$ decay, to be each greater 
than $0.5\gev$, and the transverse momentum of the slow pion from the $\dstar$
decay to be greater than $0.12\gev$;
\item $p(K \pi)/p(\pi_s) > 8$, where $p(K \pi)$ is the momentum of the 
candidate $D^0$ and $p(\pi_s)$ that of the slow pion.
\end{itemize}
Since no particle identification was performed, the $K$ and $\pi$ masses were 
alternately attributed to the decay products of the candidate $D^0$ meson.  
Only $D^0$ candidates that had an invariant mass between $1.80\gev$ and 
$1.92\gev$ were subject to the mass difference requirement 
$0.143 < \Delta M < 0.148\gev$ ($\Delta M = M(K \pi \pi_s) - M(K \pi)$).  
Further requirements placed on the $\dstar$ candidate were
$1.5 < p_T (\dstar) < 8.0\gev$ and $| \eta (\dstar) | < 1.5$.

After applying these requirements, a signal of $1720 \pm 63$ $\dstar$ 
mesons was obtained, using the fit procedure described in \Sect{sig}.

\subsection{Diffractive selection}
\label{sec-dsel}

\begin{figure}
\begin{center}
\includegraphics[height=90mm]{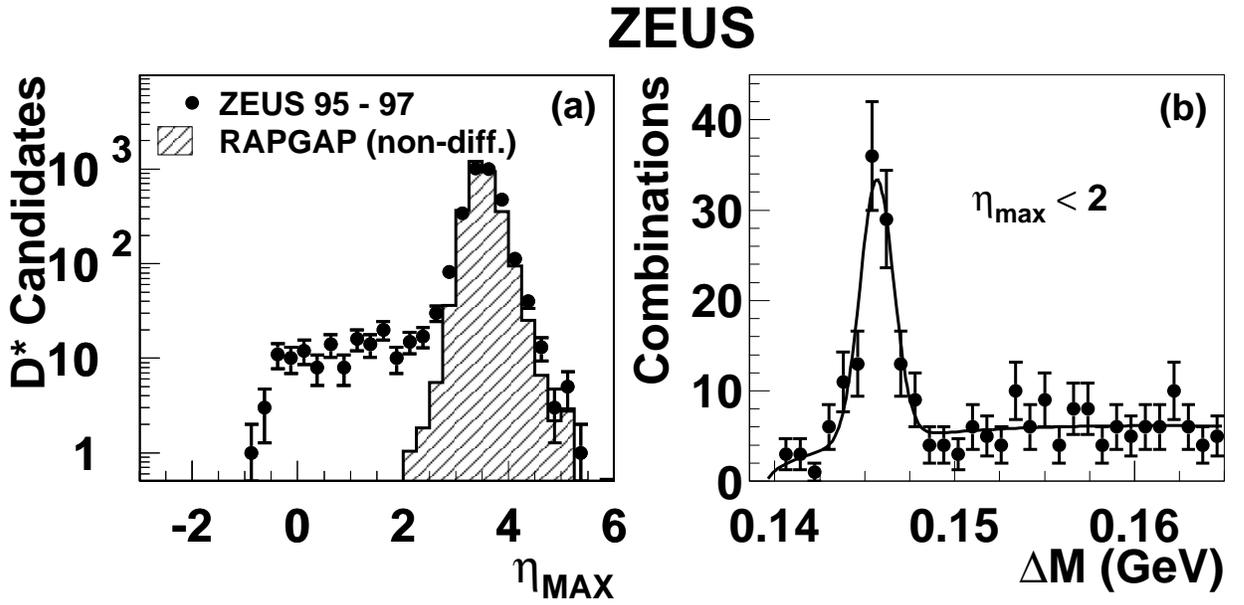}
\end{center}
\caption{(a) The $\etam$ distribution for DIS events with a $\dstar$ 
candidate.  The shaded histogram is the prediction from the RAPGAP 
non-diffractive Monte Carlo simulation (see section~5).  (b) The $\Delta M$
distribution for events with $\etam < 2$.  Only the combinations whose values
of $M(K \pi)$ lie in the signal region are included.  The curve shows 
the unbinned log-likelihood fit described in the text.}
\label{fig-mdist}
\end{figure}

A key characteristic of a diffractive event is the presence of a large 
rapidity gap between the scattered proton, which remains in the forward 
beampipe, and the hadronic system $X$ from the dissociated virtual photon.
Such events were selected using a cut on $\etam$, the 
pseudorapidity of the most-forward EFO with energy greater than $400\mev$ in 
the event~\cite{pl:b315:481,*pl:b332:228}.  \Fig{mdist}(a) shows the 
distribution of $\etam$ for events containing a $\dstar$ candidate.  The 
non-diffractive events deposit energy around the FCAL beamhole, the edge of 
which is located at a pseudorapidity of about four units, producing the peak 
at $\etam \sim 3.5$.  Below that value, the contribution from
non-diffractive interactions falls exponentially \cite{epj:c6:43}, leaving a
plateau at lower $\etam$ values, which is the signature of diffractive 
interactions.  The shaded area in \fig{mdist}(a), obtained from the 
non-diffractive Monte Carlo simulation described in Section~\ref{sec-mc}, 
exhibits an exponential fall-off at low $\etam$ and has no events with 
$\etam < 2$, confirming that the plateau observed in the data corresponds to  
diffractive events.  A cut of $\etam < 2$, i.e.~a gap of at least two units of
pseudorapidity, was used to select diffractive events.

There are two main implications of the use of the $\etam$ selection.  First, 
this method defines the rapidity gap with respect to the forward edge of the 
calorimeter and therefore cannot distinguish diffractive events in which the 
proton remains intact from those in which the proton dissociates into a 
low-mass system, whose decay products remain in the forward beampipe.  To 
correct for this, a proton-dissociative contribution of $(31 \pm 15)$\% 
\cite{epj:c6:43} has been subtracted from all measured cross sections.  
Secondly, the $\etam$ method restricts the range of $\xpom$ values; to account
for this, a cut of $\xpom < 0.016$ was applied.

The presence of charm in the diffractive events automatically sets a lower 
limit on $M_X$.  This, in turn, places an upper limit on 
the value of $\beta$ of $\beta_{max} \simeq 0.96$.  Since, in addition, the 
acceptance falls steeply at high $\beta$, only events that satisfy 
$\beta < 0.8$ were retained for further analysis.

\subsection{$\dstar$ Signal}
\label{sec-sig}

The $\Delta M$ distribution after all selection cuts is shown in 
\fig{mdist}(b).  A clear $\dstar$ signal is evident over a small background.

To determine the number of $\dstar$ candidates, the $\Delta M$ 
distribution was fitted using a Gaussian function for the signal and the form 
\begin{equation}
a (\Delta M - m_{\pi})^b e^{c (\Delta M - m_{\pi})} \nonumber
\end{equation}
for the background, where $a$, $b$ and $c$, as well as the mean, width and 
normalisation of the Gaussian function were free parameters of the fit.  The 
number of $\dstar$ candidates resulting from an unbinned log-likelihood fit was
$84 \pm 13$.  The mean value of $\Delta M$ for the peak was 
$145.67 \pm 0.14\mev$, consistent with the Particle Data Group (PDG) 
value \cite{epj:c15:1}.  The width of the Gaussian was $0.93 \pm 0.16\mev$, in 
agreement with the detector resolution.  The fit is good, as seen in 
\fig{mdist}(b).  The same fit procedure was then followed in each cross section
bin to obtain the number of $\dstar$ candidates and gave satisfactory results
in each case~\cite{thesis:cole:1999}.  The systematic uncertainties relating 
to the extraction of the number of candidates are discussed in \sect{syst}.

The $K \pi$ invariant-mass distribution was fitted with a Gaussian function
for the signal and a simple polynomial to describe the background.  The fit to
the $M(K \pi)$ distribution yielded a $D^0$ mass of $1865.7 \pm 2.1\mev$ and a 
resolution of $15 \pm 2\mev$.  The former is in good agreement with 
the PDG value \cite{epj:c15:1}, while the latter is in agreement with the 
detector resolution.  The fit results for both the $\Delta M$ and the 
$M(K \pi)$ invariant-mass distributions are comparable to those found in
the inclusive dataset~\cite{epj:c12:35,thesis:cole:1999}.

\section{Monte Carlo simulation and reweighting}
\label{sec-mc}

A GEANT-based \cite{tech:cern-dd-ee-84-1} Monte Carlo (MC) simulation was used 
to calculate selection efficiencies and correction factors.  Two different 
diffractive event generators were used:  RIDI v2.0
\cite{sovjnp:52:529,*proc:blois:1993:181,*proc:MC:1998:386} for evaluating the 
nominal correction factors and RAPGAP v2.08/01 \cite{cpc:86:147} as a 
systematic check.  For all the MC samples, events with at least one $\dstar$ 
decaying in the appropriate decay channel were selected and passed through the
standard ZEUS detector and trigger simulations as well as the event 
reconstruction package.

The RIDI generator is based on the two-gluon-exchange model developed by 
Ryskin~\cite{sovjnp:52:529,*proc:blois:1993:181,*proc:MC:1998:386}.  All stages
of the parton fragmentation and hadronisation were simulated with the Lund 
string model \cite{prep:97:31}, as implemented in JETSET \cite{cpc:82:74}.  The
simulation included QED radiative corrections.  Separate samples of 
$\gamma^* \rightarrow \qqb$ and 
$\gamma^* \rightarrow \qqb g$ interactions were generated.  The 
CTEQ4LQ~\cite{pr:d55:1280} proton parton distribution functions were used as 
input and the charm mass ($m_c$) was set to $1.35\gev$.  The acceptance 
corrections have little sensitivity to the value chosen for $m_c$.

\begin{figure}
\begin{center}
\includegraphics[height=150mm]{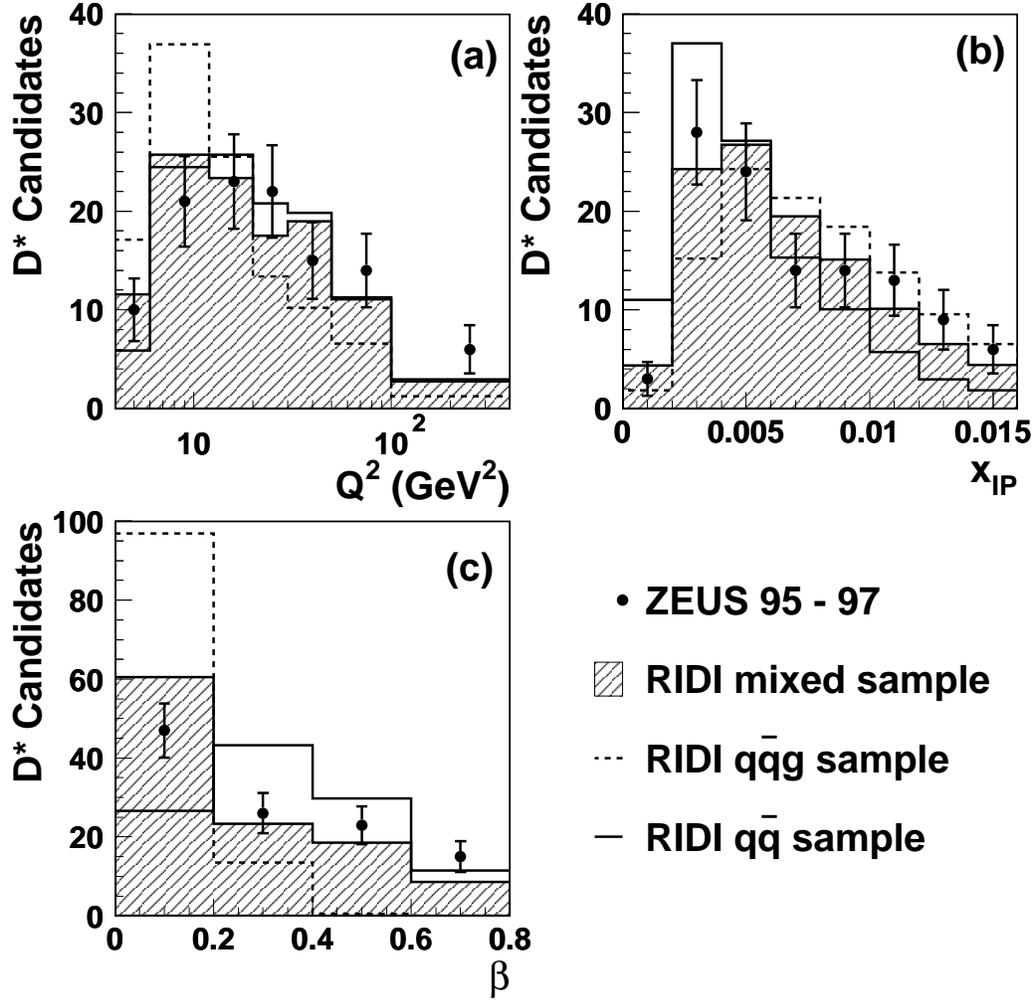}
\end{center}
\caption{The $Q^2$, $\xpom$ and $\beta$ distributions at the detector level for
the data and the RIDI MC samples.  The data sample is shown as the solid 
points and the $\qqb$ and $\qqb g$ RIDI samples are shown separately as the 
solid and dashed histograms.  The mixed sample, produced according to the 
procedure described in the text, is shown as the hatched histograms.  All the 
MC distributions have been normalised to the number of events in the data.}
\label{fig-ridi}
\end{figure}

The distributions of the variables $Q^2$, $\xpom$ and $\beta$ from the $\qqb$ 
and $\qqb g$ RIDI MC samples are compared to the data in \fig{ridi}.  Neither 
MC sample alone reproduces the measurements, particularly the $\beta$ 
distribution.  Therefore, the two RIDI MC samples were combined by 
fitting the fraction of $\qqb$ events in the three bins of $Q^2$ 
used to extract the differential cross sections.  The $\xpom$ distribution in 
each $Q^2$ bin was used in making the fit.  It was found that neither 
reweighting in $\beta$ nor in $\xpom$ separately gave a good description of 
the detector-level distributions in the data.  The fraction of $\qqb$ events 
increases with $Q^2$ from approximately $4$\% in the lowest-$Q^2$ bin to 
around $50$\% in the highest-$Q^2$ bin.  This mixed sample, shown as the 
hatched histograms in \fig{ridi}, is in satisfactory agreement with the data.  
The acceptance in the kinematic region used for the cross-section measurement 
was $17.3$\%.

The RAPGAP simulation provides a rather general framework for the generation 
of diffractive events.  The resolved-Pomeron option was used, in which charm 
quarks are produced via the leading-order BGF process.  Charm fragmentation 
was carried out using the Peterson fragmentation model with the parameter
$\epsilon_c$ set to $0.035$~\cite{pr:d27:105,*zfp:c67:27}.  The fragmentation 
process was simulated using the colour-dipole model implemented in 
ARIADNE \cite{cpc:71:15} and the hadronisation was carried out according to the
Lund string model.  The sample was generated assuming a gluon-dominated
Pomeron, with a gluon distribution peaked close to 
$\beta = 1$~\cite{pl:b428:206,zfp:c76:613}.  The Pomeron intercept was set to
$\apom (0) = 1.20$ and the charm mass was set to $1.35\gev$.  The QED 
radiative corrections are not available for diffractive charm production in 
RAPGAP.  

Similar discrepancies to those observed in the original RIDI samples were 
seen when comparing the RAPGAP MC sample to the data, with the exception that 
the $\xpom$ distribution was well described.  To correct for the 
discrepancies, the RAPGAP sample was simultaneously reweighted
in $\log{Q^2}$ and $M_X$, after which agreement was obtained in all 
distributions.

The RAPGAP generator, which also simulates non-diffractive interactions, was 
used to produce a non-diffractive $\dstar$ sample for the extraction of the 
ratio of diffractive to inclusive $\dstar$ production (see \Sect{rat}).  The
parameters used were the same as those used in the measurement of the 
inclusive DIS $\dstar$ cross sections \cite{epj:c12:35}.

\section{Systematic uncertainties on the cross section measurements}
\label{sec-syst}

The major sources of systematic uncertainties and their effect on the 
measurement of the cross section are:

\begin{itemize}
\item the selection of inclusive DIS events.  These systematic uncertainties 
were derived in the same way as for the inclusive DIS $\dstar$ analysis 
\cite{epj:c12:35} and resulted in an overall variation in the cross section of 
$^{+ 2}_{-9}$\%;
\item the selection of $\dstar$ candidates.  The minimum transverse momentum of
tracks used in the $\dstar$ reconstruction was raised and lowered by $15$\% 
and the threshold on the momentum-ratio $p(K \pi)/p(\pi_s)$ was increased 
by $0.5$ units\footnote{A trigger requirement did not permit the decrease 
of this momentum-ratio cut.}.  These variations yielded a combined uncertainty
of $\pm 6$\%;
\item the selection of diffractive events.  The $\etam$ requirement was varied 
by $\pm 0.2$ units and the EFO energy threshold applied in the reconstruction 
of $\etam$ was varied by $\pm 100\mev$.  The combined effect of these changes 
was $^{+ 8}_{- 6}$\%;
\item the model dependence of the corrections.  This uncertainty was estimated 
using the reweighted diffractive RAPGAP sample instead of the mixed RIDI 
sample.  This change resulted in a variation of $-9$\%;
\item the reweighting procedure.  The $\qqb$ fraction in each $Q^2$ bin was 
separately varied up and down by its uncertainty as determined from the fit
(see Section~\ref{sec-mc}).  The resulting deviations were summed in 
quadrature, giving a variation of \mbox{$^{+5}_{-4}$\%}. 
\end{itemize}

These systematic uncertainties were added in quadrature separately for the 
positive and negative deviations from the nominal values of the cross section 
to determine the overall systematic uncertainty of $^{+ 11}_{-16}$\%.  These 
estimates were also done in each bin in which the cross section was measured.
The sources of systematic uncertainty relating to the extraction of the number
of $\dstar$ candidates were also considered, but were found to be negligible
in comparison to those listed above.

The overall normalisation uncertainties arising from the uncertainties on 
the luminosity measurement and the $\dstar$ and $D^0$ branching fractions were
not included in the systematic uncertainty.  The uncertainty arising from the 
subtraction of proton-dissociative background, quoted separately, is 
\mbox{$\pm 22$\%}~\cite{epj:c6:43}.

\section{Results}
\subsection{Cross sections}
\label{sec-xs}

The differential cross sections in any given variable $\xi$ were determined 
using
\begin{equation}
\frac{\diff \sigma}{\diff \xi} = \frac{N_D ( 1 - f_{\rm pdiss}) }{A \cdot \Lumi \cdot B \cdot \delta \xi},
\end{equation}
where $N_D$ is the number of $\dstar$ candidates fit in a bin of width 
$\delta \xi$, $A$ is the correction factor (accounting for acceptance, 
migrations, efficiencies and radiative effects) for that bin, $\Lumi$ is 
the integrated luminosity and $B = 2.59 \pm 0.06$\% is the total 
$D^{\ast +} \rightarrow D^0 \pi^{+}_{s} \rightarrow (K^- \pi^+ ) \pi^{+}_{s}$ 
branching ratio \cite{epj:c15:1}.  The quantity $f_{\rm pdiss}$ is the fraction
of proton-dissociative background.  All other sources of background were 
neglected.

\begin{figure}
\begin{center}
\includegraphics[width=130mm]{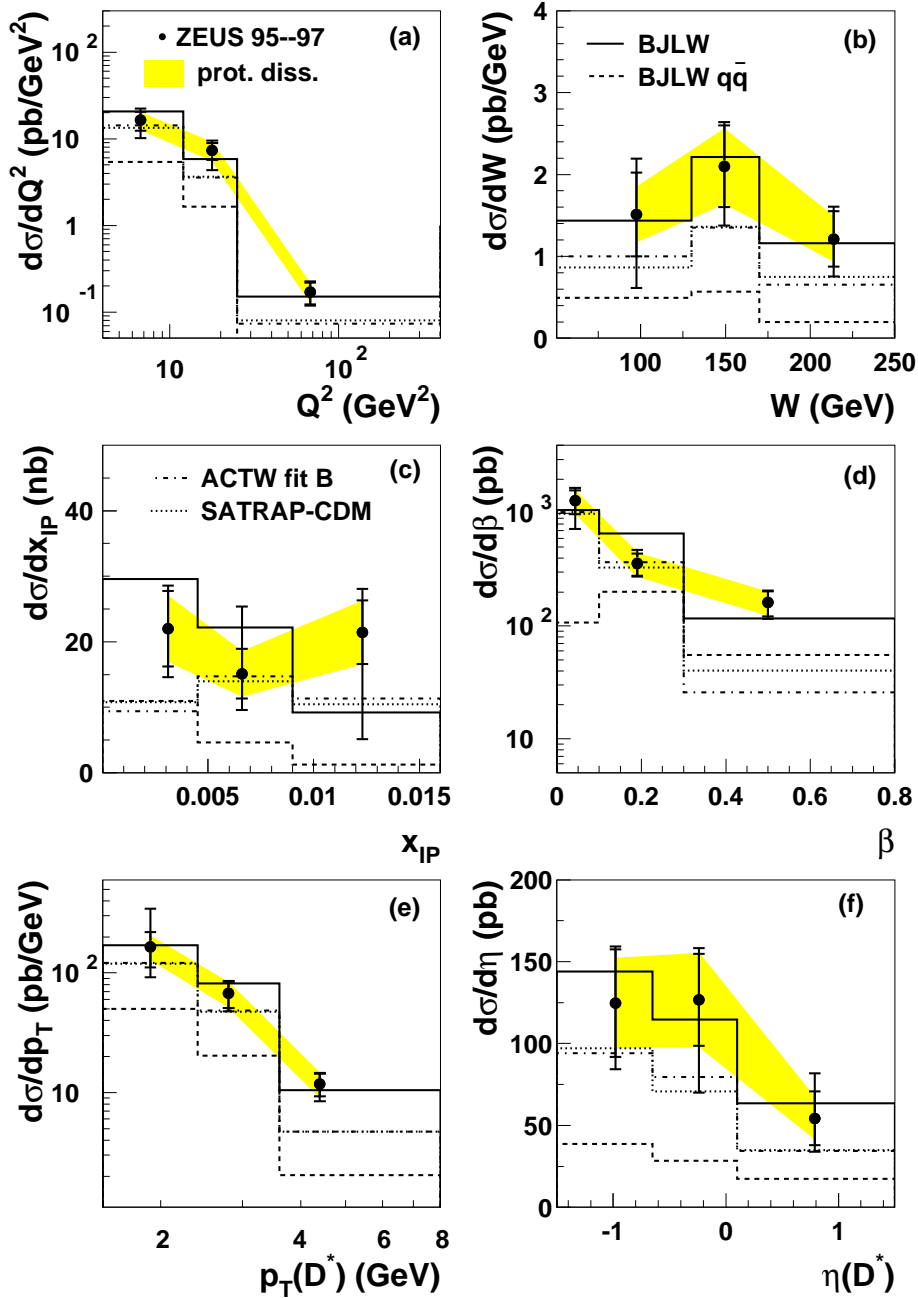}
\end{center}
\caption{Differential cross sections for $\dstar$ production in the kinematic 
region described in the text.  The cross sections are shown as a function of 
(a) $Q^2$, (b) $W$, (c) $\xpom$, (d) $\beta$, (e) $p_T (\dstar)$ and (f) 
$\eta (\dstar)$.  The inner bars show the statistical uncertainties, while 
the outer bars indicate the statistical and systematic uncertainties added in 
quadrature.  The shaded bands show the uncertainty coming from the subtraction
of the proton-dissociation background.  The histograms correspond to the 
different models described in the text.}
\label{fig-xsk2p}
\end{figure}

\Fig{xsk2p} shows the measured differential $\dstar$ cross sections with 
respect to the kinematic variables $Q^2$, $W$, $\xpom$, $\beta$, 
$p_T (\dstar)$ and $\eta (\dstar)$.  The values are also given in \tab{xsd}.
The data exhibit a sharp fall-off as a function of $Q^2$ and $p_T (\dstar)$.  
The events are concentrated at low $\beta$, but are uniformly distributed in 
$\xpom$, within the large uncertainties.  The $W$ dependence is mainly 
determined by the $p_T (\dstar)$ and $\eta (\dstar)$ restrictions.

The cross section for diffractive $\dstar$ production in the kinematic region 
$4 < Q^2 < 400\gev^2$, $0.02 < y < 0.7$, $1.5 < p_T(\dstar) < 8\gev$, 
$| \eta(\dstar)| < 1.5$, $\xpom < 0.016$ and $\beta < 0.8$ is
\begin{equation}
\sigma_{ep \rightarrow e \dstar X^{\prime} p} = 291 \pm 44 ({\rm stat.}) ^{+ 32}_{- 47} ({\rm syst.}) \pm 63 ({\rm prot.~diss.}) \pb,
\end{equation}
where the last uncertainty arises from the subtraction of the background from 
proton dissociation.

\subsection{Ratio of diffractive to inclusive $\mathbf{\dstar}$ production.}
\label{sec-rat}

The ratio of diffractive to inclusive $\dstar$ production was measured for
$x < 0.0128$.  This limit is equivalent to the $\xpom$ and $\beta$ requirements
imposed on the diffractive sample.  In the kinematic region 
$4 < Q^2 < 400\gev^2$, $0.02 < y < 0.7$, $1.5 < p_T(\dstar) < 8\gev$, 
$| \eta(\dstar)| < 1.5$ and $x < 0.0128$, the inclusive DIS $\dstar$ 
cross section is 
\mbox{$\sigma_{ep \rightarrow e \dstar Y} = 4.83 \pm 0.18 ({\rm stat.})\nb$},
where $Y$ is the complete hadronic final state except for the tagged $\dstar$ 
meson.  This value is consistent with an earlier 
ZEUS measurement\cite{epj:c12:35}, taking into account the differences between 
the two kinematic regions.  The ratio of diffractive to inclusive DIS $\dstar$ 
production is then defined by
\begin{equation}
R_D = \frac{ \sigma_{ep \rightarrow e \dstar X^{\prime} p} (\xpom < 0.016, \beta < 0.8)}{\sigma_{ep \rightarrow e \dstar Y} (x < 0.0128)}.
\end{equation}

\begin{figure}
\begin{center}
\includegraphics[height=90mm]{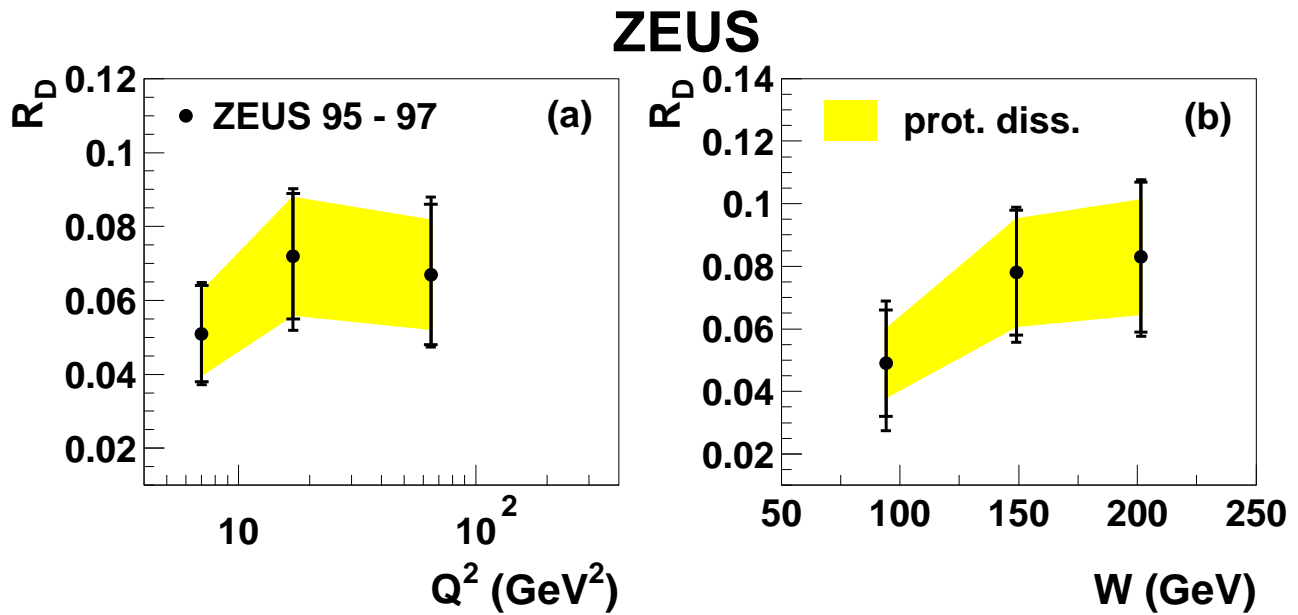}
\end{center}
\caption{The measured ratio of diffractively produced $\dstar$ mesons to 
inclusive $\dstar$ meson production, $R_D$, as a function of $Q^2$ 
and $W$.  The inner bars indicate the statistical
uncertainties, while the outer ones indicate the statistical and systematic 
uncertainties added in quadrature. The shaded band indicates the uncertainty 
arising from the subtraction of the proton-dissociation background.}
\label{fig-rk2p}
\end{figure}

It is assumed that all the systematic uncertainties, except those relating to 
the diffractive selection and the MC reweighting, cancel.  This is a reasonable
approximation given that the cross section results in the diffractive case are
statistically limited.  The measured ratio is therefore
\begin{equation}
R_D = 6.0 \pm 0.9 ({\rm stat.}) ^{+0.5}_{-0.7} ({\rm syst.}) \pm 1.3 ({\rm prot.~diss.}) \%.
\label{eq-rat}
\end{equation}
\Fig{rk2p} shows $R_D$ as a function of $Q^2$ and $W$.

\section{Discussion}

Three models are compared to the measured cross sections:  (1) the 
resolved-Pomeron model, as implemented in the fits to HERA data made by 
Alvero et al.~(ACTW) \cite{pr:d59:74022,*hep-ph-9806340}, (2) the 
two-gluon-exchange ``saturation'' model of Golec-Biernat and W\"usthoff 
\cite{pr:d59:014017,*pr:d60:114023}, as implemented in the SATRAP MC generator 
and interfaced to RAPGAP~\cite{proc:ringberg:1999:361,*proc:dis:2000:192} and 
(3) the two-gluon-exchange model of Bartels et al.~(BJLW)~\cite{pl:b379:239,*pl:b386:389,pl:b406:171,*epj:c11:111,*hep-ph-0010300}, as implemented in RAPGAP.

The version of SATRAP used is the same as that used in the recent ZEUS 
publication on the study of hadronic final states in diffraction 
\cite{pr:d65:052001}, where it is referred to as ``SATRAP-CDM''.  Although the 
SATRAP and BJLW predictions are both based on two-gluon exchange, they differ 
in the treatment of the $\qqb g$ final state, which is an important contributor
to charm production.  In the SATRAP-CDM model, only configurations in which 
the transverse momenta satisfy $k_T(q), k_T(\bar{q}) > k_T(g)$ are included in
the calculation of the $\qqb g$ final state, while all configurations with
$k_T(g) > 1\gev$ are used in the BJLW model.

The predicted cross sections from three of the fits by Alvero et al. are shown
in \tab{xs}.   The calculations were made assuming $m_c = 1.45\gev$ and the 
Peterson fragmentation model (with the parameter $\epsilon_c = 0.035$) for 
the charm decay.  The probability for charm to fragment into a $\dstar$ meson 
was taken as $0.235$ \cite{hep-ex-9912064} and the renormalisation and 
factorisation scales were set to 
$\mu_{R} = \mu_{F} = \sqrt{Q^2 + 4 m_{c}^{2}}$.  Varying the
charm mass by $\pm 0.15\gev$ gave a $^{+18}_{-17}$\% variation in each cross 
section.  The theoretical uncertainties relating to the scale and the value of
the Pomeron intercept are comparable to that arising from the charm mass.  
Fits B and D, which assume a gluon-dominated Pomeron, are favoured by the 
data.  The fit SG, which assumes a Pomeron dominated by gluons with a 
``superhard'' density, predicts too small a cross section.  A quark-dominated 
Pomeron (not shown) predicts a cross section too small by two orders of 
magnitude.

\Tab{xs} also shows the predicted cross sections from SATRAP-CDM and the BJLW
model, along with the $\qqb$ contribution from the BJLW calculations.  The 
same assumptions are made in these calculations as were made in the 
ACTW case, except that here $\mu_{R} = \mu_{F} = Q$.  The BJLW $\qqb$ 
contribution is clearly too small, while the sum of the $\qqb$ and $\qqb g$
BJLW contributions gives good agreement with the data.  The SATRAP-CDM
and ACTW fit B predictions are smaller than the data, but not significantly so.

\Fig{xsk2p} compares the predictions of ACTW fit B, SATRAP-CDM and BJLW, along 
with the $\qqb$ BJLW contribution, with the measured differential
$\dstar$ cross sections.  The ACTW fit B and fit D predict similar shapes for 
the differential distributions and therefore no comparisons with fit D are
made.  The ACTW and SATRAP-CDM predictions are in reasonable agreement with
the measured differential distributions, with the exception of the $\beta$
distribution, where both predictions undershoot the data at high $\beta$.  The
$\qqb$ contribution from the BJLW calculations clearly fails to 
describe the shape of the distributions, particularly that for $\beta$.  The
sum of the two BJLW contributions gives a good representation of all the 
measured distributions.

The ratio, $R_D$, agrees with that determined in inclusive diffraction 
\cite{epj:c6:43}, indicating that charm production is not suppressed 
in diffractive DIS, contrary to the expectations of some early 
models~\cite{zfp:c53:331}.  \Fig{rk2p} shows that $R_D$ is consistent with being
independent of $Q^2$ and $W$.

The H1 Collaboration recently published similar results  \cite{pl:b520:191}.
The kinematic region in that study is significantly different to that used 
here, making it difficult to compare the results directly.
However, comparisons with Monte Carlo models indicate that the two sets of 
results are consistent.

\section{Summary}

The cross section for diffractive $\dstar$ production in the kinematic region, 
$4 < Q^2 < 400\gev^2$, $0.02 < y < 0.7$, $\xpom < 0.016$, $\beta < 0.8$, 
$1.5 < p_T (\dstar) < 8\gev$ and $| \eta (\dstar) | < 1.5$ has been measured 
to be 
$291 \pm 44 ({\rm stat.})^{+32}_{-47}({\rm syst.}) \pm 63({\rm prot.~diss.})\pb$.  
Differential cross sections have been compared to the predictions of different
models of diffractive charm production.  The resolved-Pomeron model of Alvero
et al. is below the data at high $\beta$.  In 
the case of the two-gluon-exchange models, the $\qqb$ contribution alone gives
too small a cross section.  Inclusion of the $\qqb g$ contribution results in 
a good description of the data by the BJLW prediction, while the SATRAP-CDM 
prediction undershoots the data, particularly at high $\beta$.

The ratio of diffractive $\dstar$ production to inclusive DIS $\dstar$ 
production is 
$R_D = 6.0 \pm 0.9({\rm stat.}) ^{+0.5}_{-0.7} ({\rm syst.}) \pm 1.3({\rm prot.~diss.})$ \%.  This result is in agreement with the corresponding ratio for
inclusive diffraction, indicating that charm production in diffraction is not 
suppressed with respect to light-flavour production.  The ratio $R_D$ is  
independent of $Q^2$ and $W$ within the uncertainties.

\section*{Acknowledgments}

We thank the DESY Directorate for their strong support and encouragement, and 
the HERA machine group for their diligent efforts.  We are grateful for the 
support of the DESY computing and network services.  The design, construction
and installation of the ZEUS detector have been made possible owing to the 
ingenuity and effort of many people from DESY and home institutes who are not
listed as authors.  It is also a pleasure to thank L.~Alvero, J.~Bartels, 
J.~Collins, A.~Hebecker, H.~Jung, M.~McDermott and M.~Ryskin for useful 
discussions.

{
\def\bibname{\Large\bf References}
\def\refname{\Large\bf References}
\pagestyle{plain}
\ifzeusbst
  \bibliographystyle{./BiBTeX/bst/l4z_default}
\fi
\ifzdrftbst
  \bibliographystyle{./BiBTeX/bst/l4z_draft}
\fi
\ifzbstepj
  \bibliographystyle{./BiBTeX/bst/l4z_epj}
\fi
\ifzbstnp
  \bibliographystyle{./BiBTeX/bst/l4z_np}
\fi
\ifzbstpl
  \bibliographystyle{./BiBTeX/bst/l4z_pl}
\fi
{\raggedright
\bibliography{./BiBTeX/user/syn.bib,%
	      ./BiBTeX/user/extra.bib,%
              ./BiBTeX/bib/l4z_articles.bib,%
              ./BiBTeX/bib/l4z_books.bib,%
              ./BiBTeX/bib/l4z_conferences.bib,%
              ./BiBTeX/bib/l4z_h1.bib,%
              ./BiBTeX/bib/l4z_misc.bib,%
              ./BiBTeX/bib/l4z_old.bib,%
              ./BiBTeX/bib/l4z_preprints.bib,%
              ./BiBTeX/bib/l4z_replaced.bib,%
              ./BiBTeX/bib/l4z_temporary.bib,%
              ./BiBTeX/bib/l4z_zeus.bib}}

\providecommand{\etal}{et al.\xspace}
\providecommand{\coll}{Coll.\xspace}
\catcode`\@=11
\def\@bibitem#1{%
\ifmc@bstsupport
  \mc@iftail{#1}%
    {;\newline\ignorespaces}%
    {\ifmc@first\else.\fi\orig@bibitem{#1}}
  \mc@firstfalse
\else
  \mc@iftail{#1}%
    {\ignorespaces}%
    {\orig@bibitem{#1}}%
\fi}%
\catcode`\@=12
\begin{mcbibliography}{10}

\bibitem{zfp:c68:569}
ZEUS \coll, M.~Derrick \etal,
\newblock Z.\ Phys.{} {\bf C~68},~569~(1995)\relax
\relax
\bibitem{pl:b348:681}
H1 \coll, T.~Ahmed \etal,
\newblock Phys.\ Lett.{} {\bf B~348},~681~(1995)\relax
\relax
\bibitem{zfp:c70:391}
ZEUS \coll, M.~Derrick \etal,
\newblock Z.\ Phys.{} {\bf C~70},~391~(1996)\relax
\relax
\bibitem{epj:c6:43}
ZEUS \coll, J.~Breitweg \etal,
\newblock Eur.\ Phys.\ J.{} {\bf C~6},~43~(1999)\relax
\relax
\bibitem{pl:b428:206}
H1 \coll, C.~Adloff \etal,
\newblock Phys.\ Lett.{} {\bf B~428},~206~(1998)\relax
\relax
\bibitem{zfp:c76:613}
H1 \coll, C.~Adloff \etal,
\newblock Z.\ Phys.{} {\bf C~76},~613~(1997)\relax
\relax
\bibitem{pr:d65:052001}
ZEUS \coll, S.~Chekanov \etal,
\newblock Phys.\ Rev.{} {\bf D~65},~052001~(2002)\relax
\relax
\bibitem{np:b231:189}
A.~Donnachie and P.V.~Landshoff,
\newblock Nucl.\ Phys.{} {\bf B~231},~189~(1984)\relax
\relax
\bibitem{pl:b296:227}
A.~Donnachie and P.V.~Landshoff,
\newblock Phys.\ Lett.{} {\bf B~296},~227~(1992)\relax
\relax
\bibitem{pl:b404:353}
W.~Buchm\"uller, A.~Hebecker and M.F.~McDermott,
\newblock Phys.\ Lett.{} {\bf B~404},~353~(1997)\relax
\relax
\bibitem{epj:c1:293}
M.~Diehl,
\newblock Eur.\ Phys.\ J.{} {\bf C~1},~293~(1998)\relax
\relax
\bibitem{zfp:c74:671}
E.M.~Levin \etal,
\newblock Z.\ Phys.{} {\bf C~74},~671~(1997)\relax
\relax
\bibitem{epj:c1:547}
L.P.A.~Haakman, A.B.~Kaidalov and J.H.~Koch,
\newblock Eur.\ Phys.\ J.{} {\bf C~1},~547~(1998)\relax
\relax
\bibitem{pl:b378:347}
M.~Genovese, N.N.~Nikolaev and B.G.~Zakharov,
\newblock Phys.\ Lett.{} {\bf B~378},~347~(1996)\relax
\relax
\bibitem{pl:b406:171}
H.~Lotter,
\newblock Phys.\ Lett.{} {\bf B~406},~171~(1997)\relax
\relax
\bibitem{epj:c11:111}
J.~Bartels, H.~Jung and M.~W\"usthoff,
\newblock Eur.\ Phys.\ J.{} {\bf C~11},~111~(1999)\relax
\relax
\bibitem{hep-ph-0010300}
J.~Bartels, H.~Jung and A.~Kyrieleis,
\newblock Preprint \mbox{DESY-01-116} (\mbox{hep-ph/0010300}), 2000\relax
\relax
\bibitem{epj:c12:35}
ZEUS \coll, J.~Breitweg \etal,
\newblock Eur.\ Phys.\ J.{} {\bf C~12},~35~(2000)\relax
\relax
\bibitem{np:b545:21}
H1 \coll, C.~Adloff \etal,
\newblock Nucl.\ Phys.{} {\bf B~545},~21~(1999)\relax
\relax
\bibitem{pl:b152:256}
G.~Ingelman and P.E.~Schlein,
\newblock Phys.\ Lett.{} {\bf B~152},~256~(1985)\relax
\relax
\bibitem{sovjnp:15:438}
V.N.~Gribov and L.N.~Lipatov,
\newblock Sov.\ J.\ Nucl.\ Phys.{} {\bf 15},~438~(1972)\relax
\relax
\bibitem{sovjnp:20:94}
L.N.~Lipatov,
\newblock Sov.\ J.\ Nucl.\ Phys.{} {\bf 20},~94~(1975)\relax
\relax
\bibitem{jetp:46:641}
Yu.L.~Dokshitzer,
\newblock Sov.\ Phys.\ JETP{} {\bf 46},~641~(1977)\relax
\relax
\bibitem{np:b126:298}
G.~Altarelli and G.~Parisi,
\newblock Nucl.\ Phys.{} {\bf B~126},~298~(1977)\relax
\relax
\bibitem{pl:b356:129}
ZEUS \coll, M.~Derrick \etal,
\newblock Phys.\ Lett.{} {\bf B~356},~129~(1995)\relax
\relax
\bibitem{QCDmodels}
See e.g.,
\newblock {\em Proc.\ {HERA} Workshop}, G.~Ingelman, A.~De Roeck and
  R.~Klanner~(eds.), Vol.~2, p.~635.
\newblock DESY (1996).
\newblock See also references therein\relax
\relax
\bibitem{pr:d12:163}
F.E.~Low,
\newblock Phys.\ Rev.{} {\bf D~12},~163~(1975)\relax
\relax
\bibitem{prl:34:1286}
S.~Nussinov,
\newblock Phys.\ Rev.\ Lett.{} {\bf 34},~1286~(1975)\relax
\relax
\bibitem{pr:d14:246}
S.~Nussinov,
\newblock Phys.\ Rev.{} {\bf D~14},~246~(1976)\relax
\relax
\bibitem{zfp:c53:331}
N.N.~Nikolaev and B.G.~Zakharov,
\newblock Z.\ Phys.{} {\bf C~53},~331~(1992)\relax
\relax
\bibitem{jetp:81:625}
M.~Genovese, N.N.~Nikolaev and B.G.~Zakharov,
\newblock Sov.\ Phys.\ JETP{} {\bf 81},~625~(1995)\relax
\relax
\bibitem{pl:b366:371}
A.~Edin, G.~Ingelman and J.~Rathsman,
\newblock Phys.\ Lett.{} {\bf B~366},~371~(1996)\relax
\relax
\bibitem{pl:b520:191}
H1 \coll, C.~Adloff \etal,
\newblock Phys.\ Lett.{} {\bf B~520},~191~(2001)\relax
\relax
\bibitem{zeus:1993:bluebook}
ZEUS \coll, U.~Holm~(ed.),
\newblock {\em The {ZEUS} Detector}.
\newblock Status Report (unpublished), DESY (1993),
\newblock available on
  \texttt{http://www-zeus.desy.de/bluebook/bluebook.html}\relax
\relax
\bibitem{nim:a279:290}
N.~Harnew \etal,
\newblock Nucl.\ Inst.\ Meth.{} {\bf A~279},~290~(1989)\relax
\relax
\bibitem{npps:b32:181}
B.~Foster \etal,
\newblock Nucl.\ Phys.\ Proc.\ Suppl.{} {\bf B~32},~181~(1993)\relax
\relax
\bibitem{nim:a338:254}
B.~Foster \etal,
\newblock Nucl.\ Inst.\ Meth.{} {\bf A~338},~254~(1994)\relax
\relax
\bibitem{nim:a309:77}
M.~Derrick \etal,
\newblock Nucl.\ Inst.\ Meth.{} {\bf A~309},~77~(1991)\relax
\relax
\bibitem{nim:a309:101}
A.~Andresen \etal,
\newblock Nucl.\ Inst.\ Meth.{} {\bf A~309},~101~(1991)\relax
\relax
\bibitem{nim:a321:356}
A.~Caldwell \etal,
\newblock Nucl.\ Inst.\ Meth.{} {\bf A~321},~356~(1992)\relax
\relax
\bibitem{nim:a336:23}
A.~Bernstein \etal,
\newblock Nucl.\ Inst.\ Meth.{} {\bf A~336},~23~(1993)\relax
\relax
\bibitem{nim:a401:63}
A.~Bamberger \etal,
\newblock Nucl.\ Inst.\ Meth.{} {\bf A~401},~63~(1997)\relax
\relax
\bibitem{nim:a382:419}
A.~Bamberger \etal,
\newblock Nucl.\ Inst.\ Meth.{} {\bf A~382},~419~(1996)\relax
\relax
\bibitem{desy-92-066}
J.~Andruszk\'ow \etal,
\newblock Preprint \mbox{DESY-92-066}, DESY, 1992\relax
\relax
\bibitem{zfp:c63:391}
ZEUS \coll, M.~Derrick \etal,
\newblock Z.\ Phys.{} {\bf C~63},~391~(1994)\relax
\relax
\bibitem{acpp:b32:2025}
J.~Andruszk\'ow \etal,
\newblock Acta Phys.\ Pol.{} {\bf B~32},~2025~(2001)\relax
\relax
\bibitem{nim:a365:508}
H.~Abramowicz, A.~Caldwell and R.~Sinkus,
\newblock Nucl.\ Inst.\ Meth.{} {\bf A~365},~508~(1995)\relax
\relax
\bibitem{briskin:phd:1998}
G.M.~Briskin,
\newblock {\em Diffractive Dissociation in $ep$ Deep Inelastic Scattering}.
\newblock Ph.D.\ Thesis, Tel Aviv University, 1998.
\newblock (Unpublished)\relax
\relax
\bibitem{epj:c1:81}
ZEUS \coll, J.~Breitweg \etal,
\newblock Eur.\ Phys.\ J.{} {\bf C~1},~81~(1998)\relax
\relax
\bibitem{proc:hera:1991:23}
S.~Bentvelsen, J.~Engelen and P.~Kooijman,
\newblock {\em Proc.\ Workshop on Physics at {HERA}}, W.~Buchm\"uller and
  G.~Ingelman~(eds.), Vol.~1, p.~23.
\newblock Hamburg, Germany, DESY (1992)\relax
\relax
\bibitem{proc:hera:1991:43}
K.C.~H\"oger,
\newblock {\em Proc.\ Workshop on Physics at {HERA}}, W.~Buchm\"uller and
  G.~Ingelman~(eds.), Vol.~1, p.~43.
\newblock Hamburg, Germany, DESY (1992)\relax
\relax
\bibitem{prl:35:1672}
S.~Nussinov,
\newblock Phys.\ Rev.\ Lett.{} {\bf 35},~1672~(1975)\relax
\relax
\bibitem{pl:b315:481}
ZEUS \coll, M.~Derrick \etal,
\newblock Phys.\ Lett.{} {\bf B~315},~481~(1993)\relax
\relax
\bibitem{pl:b332:228}
ZEUS \coll, M.~Derrick \etal,
\newblock Phys.\ Lett.{} {\bf B~332},~228~(1994)\relax
\relax
\bibitem{epj:c15:1}
Particle Data Group, D.E. Groom \etal,
\newblock Eur.\ Phys.\ J.{} {\bf C~15},~1~(2000)\relax
\relax
\bibitem{thesis:cole:1999}
J.E.~Cole,
\newblock {\em Open Charm Production in Deep Inelastic Diffractive $ep$
  Scattering at HERA}.
\newblock Thesis, University of London, Report \mbox{RAL-TH-1999-008},
  1999\relax
\relax
\bibitem{tech:cern-dd-ee-84-1}
R.~Brun et al.,
\newblock {\em {\sc geant3}},
\newblock Technical Report CERN-DD/EE/84-1, CERN, 1987\relax
\relax
\bibitem{sovjnp:52:529}
M.G.~Ryskin,
\newblock Sov.\ J.\ Nucl.\ Phys.{} {\bf 52},~529~(1990)\relax
\relax
\bibitem{proc:blois:1993:181}
M.G.~Ryskin, S.Y.~Sivoklokov and A.~Solano,
\newblock {\em Proc.\ International Conf.\ on Elastic and Diffractive
  Scattering, Providence RI, June 1993}, H.M. Fried, K.~Kang and
  C.I.~Tan~(eds.).
\newblock World Scientific, Singapore (1993)\relax
\relax
\bibitem{proc:MC:1998:386}
M.G.~Ryskin and A.~Solano,
\newblock {\em Proc.\ Workshop on Monte Carlo Generators for {HERA} Physics},
  G.~Grindhammer, G.~Ingelman, H.~Jung and T.~Doyle~(eds.), p.~386.
\newblock DESY, Hamburg, Germany (1999).
\newblock Also in preprint \mbox{DESY-PROC-1999-02},
\newblock available on \texttt{http://www.desy.de/\til heramc/}\relax
\relax
\bibitem{cpc:86:147}
H.~Jung,
\newblock Comp.\ Phys.\ Comm.{} {\bf 86},~147~(1995)\relax
\relax
\bibitem{prep:97:31}
B.~Andersson \etal,
\newblock Phys.\ Rep.{} {\bf 97},~31~(1983)\relax
\relax
\bibitem{cpc:82:74}
T.~Sj\"ostrand,
\newblock Comp.\ Phys.\ Comm.{} {\bf 82},~74~(1994)\relax
\relax
\bibitem{pr:d55:1280}
H.L.~Lai \etal,
\newblock Phys.\ Rev.{} {\bf D~55},~1280~(1997)\relax
\relax
\bibitem{pr:d27:105}
C.~Peterson \etal,
\newblock Phys.\ Rev.{} {\bf D~27},~105~(1983)\relax
\relax
\bibitem{zfp:c67:27}
OPAL \coll, R.~Akers \etal,
\newblock Z.\ Phys.{} {\bf C~67},~27~(1995)\relax
\relax
\bibitem{cpc:71:15}
L.~L\"onnblad,
\newblock Comp.\ Phys.\ Comm.{} {\bf 71},~15~(1992)\relax
\relax
\bibitem{pr:d59:74022}
L.~Alvero \etal,
\newblock Phys.\ Rev.{} {\bf D~59},~074022~(1999)\relax
\relax
\bibitem{hep-ph-9806340}
L.~Alvero, J.C.~Collins and J.J.~Whitmore,
\newblock Preprint \mbox{hep-ph/9806340}, 1998\relax
\relax
\bibitem{pr:d59:014017}
K.~Golec-Biernat and M.~W\"usthoff,
\newblock Phys.\ Rev.{} {\bf D~59},~014017~(1999)\relax
\relax
\bibitem{pr:d60:114023}
K.~Golec-Biernat and M.~W\"usthoff,
\newblock Phys.\ Rev.{} {\bf D~60},~114023~(1999)\relax
\relax
\bibitem{proc:ringberg:1999:361}
H.~Kowalski,
\newblock {\em Proceedings of the Workshop on New Trends in HERA Physics},
  G.~Grindhammer, B.A.~Kniehl and G.~Kramer~(eds.), pp.~361--380.
\newblock  (1999),
\newblock available on
  \texttt{http://www-library.desy.de/conf/ringberg99.html}\relax
\relax
\bibitem{proc:dis:2000:192}
H.~Kowalski and M.~W\"usthoff,
\newblock {\em Proceedings of the 8th International Workshop Deep Inelastic
  Scattering and QCD}, J.~Gracey and T.~Greenshaw~(eds.), p.~192.
\newblock World Scientific, Singapore (2000)\relax
\relax
\bibitem{pl:b379:239}
J.~Bartels, H.~Lotter and M.~W\"usthoff,
\newblock Phys.\ Lett.{} {\bf B~379},~239~(1996)\relax
\relax
\bibitem{pl:b386:389}
J.~Bartels \etal,
\newblock Phys.\ Lett.{} {\bf B~386},~389~(1996)\relax
\relax
\bibitem{hep-ex-9912064}
L.~Gladilin,
\newblock Preprint \mbox{hep-ex/9912064}, 1999\relax
\relax
\end{mcbibliography}
}
\vfill\eject

%
%
\begin{table}[hbt]
\begin{center}
\begin{tabular}{|l|c|c|} \hline
\multicolumn{3}{|c|}{$\diff \sigma / \diff Q^2$} \\ \hline
$Q^2$ Range ($\gev^2$) & $Q^2$ ($\gev^2$) & $\sigma$ (pb) \\ \hline
$4$ - $12$   & $6.7$  & $131 \pm 32 _{-38}^{+35} \pm 28$ \\
$12$ - $25$  & $17.8$ & $95 \pm 21 _{-32}^{+19} \pm 21$  \\
$25$ - $400$ & $67.7$ & $64 \pm 18 _{-7}^{+10} \pm 14$   \\ \hline
\multicolumn{3}{|c|}{$\diff \sigma / \diff W$} \\ \hline
$W$ Range ($\gev$) & $W$ ($\gev$) & $\sigma$ (pb) \\ \hline
$50$ - $130$  & $97.4$  & $121 \pm 41 _{-59}^{+36} \pm 26$ \\
$130$ - $170$ & $149.3$ & $84 \pm 20 _{-21}^{+8} \pm 18$   \\
$170$ - $250$ & $213.9$ & $97 \pm 27 _{-25}^{+16} \pm 21$  \\ \hline
\multicolumn{3}{|c|}{$\diff \sigma / \diff \xpom$} \\ \hline
$\xpom$ Range & $\xpom$ & $\sigma$ (pb) \\ \hline
$0$ - $0.0045$     & $0.0031$ & $99 \pm 26 _{-21}^{+14} \pm 22$   \\
$0.0045$ - $0.009$ & $0.0066$ & $68 \pm 17 _{-18}^{+43} \pm 15$   \\
$0.009$ - $0.016$  & $0.0123$ & $150 \pm 34 _{-109}^{+32} \pm 33$ \\ \hline
\multicolumn{3}{|c|}{$\diff \sigma / \diff \beta$} \\ \hline
$\beta$ Range & $\beta$ & $\sigma$ (pb) \\ \hline
$0$ - $0.1$   & $0.043$ & $129 \pm 31 _{-47}^{+22} \pm 28$ \\
$0.1$ - $0.3$ & $0.19$  & $72 \pm 16 _{-6}^{+16} \pm 16$   \\
$0.3$ - $0.8$ & $0.50$  & $81 \pm 20 _{-12}^{+9} \pm 18$   \\ \hline
\multicolumn{3}{|c|}{$\diff \sigma / \diff p_T (\dstar)$} \\ \hline
$p_T(\dstar)$ Range ($\gev$) & $p_T(\dstar)$ ($\gev$) & $\sigma$ (pb) \\ \hline
$1.5$ - $2.4$ & $1.9$ & $149 \pm 49 _{-45}^{+152} \pm 32$ \\
$2.4$ - $3.6$ & $2.8$ & $81 \pm 20 _{-13}^{+9} \pm 18$    \\
$3.6$ - $8.0$ & $4.4$ & $52 \pm 11 _{-10}^{+5} \pm 11$    \\ \hline
\multicolumn{3}{|c|}{$\diff \sigma / \diff \eta (\dstar)$} \\ \hline
$\eta(\dstar)$ Range & $\eta(\dstar)$ & $\sigma$ (pb) \\ \hline
$-1.5$ - $-0.65$ & $-0.98$ & $106 \pm 28 _{-20}^{+9} \pm 23$ \\
$-0.65$ - $0.1$  & $-0.24$ & $95 \pm 21 _{-37}^{+11} \pm 21$ \\
$0.1$ - $1.5$    & $0.79$  & $76 \pm 23 _{-17}^{+31} \pm 17$ \\ \hline
\end{tabular}
\caption{Values of the differential cross sections with respect to $Q^2$, $W$,
$\xpom$, $\beta$, $p_T(\dstar)$ and $\eta(\dstar)$.  The following
quantities are given:  the range of the measurement; the value at which the 
cross section is quoted and the measured cross section.  The first uncertainty
quoted represents the statistical uncertainty, the second the systematic and 
the third the uncertainty arising from the subtraction of the proton 
dissociation background.}
\label{tab-xsd}
\end{center}
\end{table}

\begin{table}[hbt]
\begin{center}
\begin{tabular}{|l|c|} \hline
                      & Cross Section (pb)     \\ \hline
Data                  & $291 \pm 44 ^{+ 32}_{- 47} \pm 63$ \\ \hline
ACTW FIT B            & $187$         \\
ACTW FIT D            & $401$         \\
ACTW FIT SG           & $87$          \\ \hline
SATRAP                & $185$         \\
BJLW $\qqb$           & $79$          \\
BJLW $\qqb + \qqb g$  & $297$         \\ \hline
\end{tabular}
\caption{Comparison of the measured cross section with those predicted by the 
models described in the text.  No errors are quoted given the considerable
flexibility in the predicted cross sections depending upon the choice of 
input parameters for the models.}
\label{tab-xs}
\end{center}
\end{table}

%
%
\end{document}